\documentstyle{elsart}
\begin{document}
\begin{frontmatter}
\title{La Palma night-sky brightness}
\author{C.R. Benn}
\address{Isaac Newton Group, Apartado 321, 38780 Santa Cruz de La Palma, Spain}
\author{S.L. Ellison\thanksref{SLE}}
\address{Department of Physics, University of Kent, Canterbury, Kent, UK}
\thanks[SLE]{Now at Institute of Astronomy, Madingley Road, Cambridge,
CB3 OHA, UK}

\begin{abstract}

The brightness of the moonless night sky above La Palma 
was measured on 427 CCD images taken with the
Isaac Newton and Jacobus Kapteyn Telescopes on 63 nights during 1987 - 1996.
The median sky brightness at high elevation, high galactic latitude 
and high ecliptic latitude, at sunspot minimum,
is B = 22.7, V = 21.9, R = 21.0,
similar to that at other dark sites.
The main contributions to sky brightness are
airglow and zodiacal light.
The sky is brighter at low
ecliptic latitude (by 0.4 mag); at solar maximum (by 0.4 mag);
and at high airmass (0.25 mag brighter at airmass 1.5).
Light pollution 
(line + continuum) contributes $<$ 0.03 mag in $U$,
$\approx$ 0.02 mag in $B$, $\approx$ 0.10 mag in $V$, $\approx$ 0.10 mag in
$R$ at the zenith.

This paper is a summary of results which are presented in full
elsewhere (Benn \& Ellison 1998).
\end{abstract}
\end{frontmatter}

\section{Introduction}
The zenith brightness of the moonless night sky at a clear dark observing
site, measured at high ecliptic and galactic latitudes, and during
solar minimum, is typically 
B = 22.9 mag arcsec$^{-2}$, V = 21.9 mag arcsec$^{-2}$,
about 10 million times 
dimmer than the daylight sky (but easily visible
to the dark-adapted eye).
This glow\footnote{1 
$S_{10} \equiv$ one 10-th mag star per 
square degree, 220 $S_{10} \equiv$    $V$ = 21.9 mag arcsec$^{-2}$} comes from 
airglow (145 $S_{10}$),
zodiacal light (60 $S_{10}$), the integrated light of faint stars ($<$ 
5 $S_{10}$),
starlight scattered by interstellar dust (10 $S_{10}$), 
and extragalactic light ($\sim$ 1 $S_{10}$).
Auroral light is significant at
geomagnetic latitude $>$ 40$^o$.

\begin{figure}[htbp]
\vspace{80mm}
\centerline{\hbox{
}}
\caption[]{\small
Typical spectrum of the La Palma sky
on a moonless night.
Most of the distinctive features of the night-sky spectrum are due to airglow.
The NaD emission at 5890/6 \AA\ is partly from streetlighting, 
the mercury emission at 4358, 5461 \AA\ wholly so.
With the exception  of the 8645-\AA\ O$_2$ line,
the features dominating the spectrum redward of 6500 \AA\ are 
the Meinel rotation-vibration bands of OH.
}
\end{figure}

The airglow is emitted by atoms and molecules in the upper atmosphere
which are excited by solar UV radiation during the day.  
Its intensity
correlates with solar activity, being
$\approx$ 0.5 mag brighter at solar maximum, and
can also vary randomly, by up to a few
10s of \%,  with position on the sky and
with time during the night.
The strength of at least the NaD line varies with season; $\sim$ 30 Rayleighs
in local summer vs $\sim$ 180 Rayleighs in winter. 

Zodiacal light is sunlight scattered by interplanetary dust.
At high ecliptic latitude, it contributes $\approx$ 60 $S_{10}$.
Its brightness rises slowly with decreasing ecliptic latitude, to
about 140 $S_{10}$ on the ecliptic plane,
for ecliptic longitde $>$ 90$^{o}$ from the sun.
The spectrum of zodiacial light is very similar to that of the sun 
over the UV - IR range,
and it's fractional contribution to the brightness of the night sky peaks
at a wavelength of 4500 \AA\ ($\approx$ 0.5 of total 
for $\beta = 30^o$).
 
Starlight contributes substantially to the integrated brightness of the
sky, $\approx 25 + 250 e^{-(|b|/20^o)}$ $S_{10}$ units
(approximate fit to the data of Roach \& Gordon 1973,
$b$ = galactic latitude).
If all the starlight were scattered uniformly over the sky, it would produce
a background of $\approx$ 100 $S_{10}$ units.
However, most of this light is from stars with 6 $< V <$ 16 and 
published
measurements of sky brightness
usually now refer to the sky {\it between} stars with $V <$ 20.
Starlight scattered by interstellar dust produces a diffuse glow
concentrated along the galactic plane, analagous to the zodiacal light
along the ecliptic.
Faint galaxies contribute
$<$ 5 S$_{10}$ (observational upper limit) to the brightness of the night sky; 
a lower limit can be estimated from the faint galaxy counts,
$>\sim$ 1 $S_{10}$.

Light pollution at observatory sites
arises principally 
from tropospheric scattering of light emitted by sodium- and mercury-vapour
and incandescent street lamps.\footnote{Lighting wastefully emitted 
above the horizontal costs the
US taxpayer $\sim$ $\$$10$^9$ year$^{-1}$, more than the cost
of funding US astronomy (Hunter \& Crawford 1991).}
Garstang (1989) estimates
the increase in brightness at zenith
distance 45$^o$, in the direction of a 
conurbation of population P at distance D km to be
$\sim PD^{-2.5}$/70 mag.
The IAU recommendations for a dark site are continuum $\Delta mag <$ 0.1 for
3000 $< \lambda <$ 10000 \AA, and intensity of NaD light pollution
$<$ that from airglow,
at $ZD$ = 45$^o$ towards any city (Smith 1979).

At high airmass, light from outside the atmosphere may be dimmed
by scattering and absorption, 
but the 
airglow will probably be brighter, since a line of sight intercepts a larger
number of atoms in the airglow layer.  
For airglow arising in a thin layer at height $h$,
the intensity should vary 
with zenith distance $ZD$ as
$(1-(a/(a+h)^2sin^2ZD))^{-0.5} $
where $a$ is the radius of the earth
(the van Rhijn formula).

\section{Measurements from La Palma}
The observatory on La Palma 
lies close to the island's peak, on the rim of a large
volcanic caldera, at longitude 18$^o$ W, 
latitude 29$^o$ N, altitude 2300 m,
geomagnetic latitude $\approx$ 20$^o$ N. 
Approximately 70\% of the nights are clear
(95\% in the summer) and 60\% are photometric.
The median site seeing is $\approx$ 0.7 arcsec.
Atmospheric  
extinction is typically 0.15 mag in $V$, but can be substantially higher
during the summer, when dust from the Sahara desert 
(400 km away) blows over the Canary
Islands.  
The island has a population of $\approx$ 80000 people,
but 
light-pollution is strictly controlled,
by a 1992 decree, the `Ley del Cielo'.

Dark-of-moon sky brightness was measured from 427 CCD images 
taken 
with the Isaac Newton and Jacobus Kapteyn Telescopes on
63 photometric nights.
The measurements were made in areas
free of nebulosity, stars, cosmic-ray events or CCD defects. 
The median values of sky brightness
during solar minimum 1994-6
are $B$ = 22.7 $\pm$ 0.03, $V$ = 21.9 $\pm$ 0.03, $R$ =
21.0 $\pm$ 0.03 
mag arcsec$^{-2}$, with a scatter of about 0.15 mag in each band.
We have fewer measurements in $U$ and $I$ bands; the median sky brightnesses
are $U \approx$ 22.0, $I \approx$ 20.0 mag arcsec$^{-2}$,
consistent with the few values measured elsewhere.
The $I$ brightness is dominated by OH emission bands (Fig. 1)
and varies
by up to a factor $\sim$ 2 during the night.
The brightnesses in $V$ and $R$ 
are similar to those measured at other sites.
The measured brightness in $B$ is 0.2 mag higher than the sunspot-minimum
value reported for other sites at previous minima, 
but the La Palma median is dominated by data from 1995, and 
Krisciunas (1997) found that the $B$ sky brightness at Mauna Kea was
still declining then.
The 0.15-mag scatter in the measured values
is a combination of measurement errors and true variations
in sky brightness.

\begin{figure}[htbp]
\vspace{-0.5cm}
\centerline{\hbox{
}}
\vspace{80mm}
\caption[]{
\small La Palma combined $B-0.8, V$ sky brightness vs date.
Stars: ecliptic latitude $\beta >$ 40 deg;
galactic latitude $>$ 10 deg;
airmass $<$ 1.4;
mean V extinction during the night $<$ 0.2 mag.
Circles: same criteria except that
$\beta <$ 40$^o$.
In order to reduce the scatter, the measured sky brightnesses 
at low ecliptic latitude have been
corrected for the excess contribution from
zodiacal light.
The curve indicates the expected
variation of sky brightness with date, using
the $S_{10.7cm}$ flux density as an indicator
of solar activity.
}
\end{figure}
The measured sky brightness varies with 
solar cycle, being
0.4 $\pm$ 0.1 mag brighter in 1990 than in 1995 (Fig. 2).
It also varies with 
ecliptic latitude, being
0.4 mag brighter on the ecliptic than
at the poles, consistent with the known variation in the brightness
of the zodiacal light.
The sky brightens steadily with rising airmass
of observation, being 
0.25 +- 0.07 mag brighter at airmass $\sim$ 1.5 than it is
at the zenith, consistent with the eqution at the end of Section 1, if airglow contributes 70\% of the total.  
The sky is brighter under exceptionally dusty conditions, with
extincton $A_V >$ 0.25 mag.

80000 people live on La Palma, mostly in or near
nine small towns lying between 10
and 15 km from the observatory.
The predicted brightening of the zenith sky from the 
main populated areas, including
the neighbouring islands
of La Gomera, El Hierro and Tenerife,
is given in the last column of the table overleaf,
using the model of Garstang (1989).
The predictions are approximate, but serve to indicate the
relative sizes of the light-pollution contributions expected from
different sources.

\begin{table}[htbp]
\begin{tabular}{lrrrl}
Place           & Azimuth & Population& Distance & $\Delta mag_G$   \\
                &$^o$     & (1986)    & km       &               \\
Barlovento \rule{0mm}{5mm}
                &  50     &   3000    &   10     & 0.06  \\
Los Sauces / San Andres 
                & 110     &   6000    &   12     & 0.08  \\
Tenerife        & 115     & 600000    &  130     & 0.02 \\
Santa Cruz      & 125     &  18000    &   15     & 0.18 \\
Brena Alta/Baja & 140     &   8000    &   15     & 0.08 \\
Gomera          & 140     &  17000    &   90     & $<$0.01  \\
El Paso         & 180     &   7000    &   12     & 0.10  \\
Hierro          & 190     &   7000    &  110     & $<$0.01 \\
Los Llanos / Tazacorte 
                & 200     &  23000    &   12     & 0.32  \\
\end{tabular}
\end{table}

Most of the expected contamination is from La Palma's 
14000 streetlamps, which emit $\approx$ 120 Mlumens, 
corresponding to $\sim$ 180 kW of visible photons.
Approximately half of the light generated emerges from the lamp housings,
and $\approx$ 10\% of the emerging light is reflected from the ground,
so $\approx$ 9 kW
escapes to the sky.
The $V$-band zenith sky brightness on La Palma is similar to that at
other dark sites, which suggests that light pollution must contribute
$\sim<$ 0.1 mag to the total.
A stronger limit on the
contribution of light pollution to the brightness of the sky has been obtained
from the equivalent widths of the NaD and Hg lines
in the spectrum of the zenith night sky.
These yield directly the contribution to sky brightness from emission lines,
while that from 
broad and continuum features can be inferred from the strengths of
the 
 emission lines, the known shapes of the lamp spectra,
and the relative numbers of lamps of different kinds.
>From these, we estimate that 
light pollution contributes $<$ 0.03 mag to the zenith
{\it continuum}
sky brightness in 
all bands;
and that
the {\it total} contamination is $<$ 0.03 mag in $U$ (Hg lines), 
$\approx$ 0.02 mag in $B$ (Hg lines),
$\approx$ 0.10 mag in $V$ (mainly low-pressure and high-pressure NaD), 
$\approx$ 0.10 mag in $R$ (NaD, as for V band).
This degree of light pollution is comparable to that at Kitt Peak,
$\sim$
0.02 mag in $B$, 0.05 mag in $V$ in 1988 (Massey {\it{\it et al}} 1990)
from Tucson (275000 inhabitants, 65 km distant).
It is likely to {\it decrease} in the future.

\section{Conclusions}
$\bullet$
The brightness of the moonless night sky above La Palma, at high ecliptic
and galactic latitudes and low airmass, at solar minimum,
is given in various units in columns 5 - 9 of the table overleaf.

\begin{table}[htbp]
\begin{tabular}{rrrrrrrrr}
&&log10&$Jy$&mag&&$\mu$Jy&&
 \\
Band&$\lambda/\mu$      
& ($\nu$/Hz)&mag=0&arcsec$^{-2}$&$S_{10}$&arcsec$^{-2}$
&$\gamma$&R/\AA\ \\
U&  0.36& 14.92&   1810&  22.0&       205&     2.87& 0.012&    0.64\\
B&  0.44& 14.83&   4260&  22.7&       107&     3.54& 0.012&    0.65\\
V&  0.55& 14.74&   3640&  21.9&       224&     6.33& 0.017&    0.93\\
R&  0.64& 14.67&   3080&  21.0&       515&    12.26& 0.029&    1.55\\
I&  0.79& 14.58&   2550&  20.0&      1294&    25.50& 0.049&    2.61\\
\end{tabular}
\end{table}

Column 4 gives
the equivalent intensity in Jy for apparent mag = 0.
The units of column 8 
($\gamma$) are photons s$^{-1}$m$^{-2}$\AA$^{-1}$arcsec$^{-2}$. 
The brightness of the 
La Palma sky is similar to that at other dark sites.
\\$\bullet$ The sky is $\approx$ 0.4 mag brighter on the ecliptic plane 
than at the ecliptic pole,
$\approx$ 0.4 mag brighter at solar maximum
than at solar minimum, and
$\approx$ 0.25 mag brighter at zenith distance
45$^o$ than at the zenith.
\\$\bullet$ 
Light pollution contributes $<$ 0.03 mag to the zenith continuum 
sky brightness in all bands, well below the 0.1-mag limit recommended by
IAU for a dark site.
The total contamination is $<$ 0.03 mag in $U$, $\approx$ 0.02 mag in $B$,
$\approx$ 0.10 mag in $V$, $\approx$ 0.10 mag in $R$.

For sky-limited exposures,
an uncertainty of 0.4 mag in the surface brightness of the night sky translates
into an uncertainty of a factor of 2 in the exposure time required to reach
a given signal-to-noise, with potential wastage of half that time.
The measurements reported here 
of the absolute level of sky brightness, and of 
the ways in which it varies, allow more efficient use to be made of
telescope time.  

We are grateful to 
Ed Zuiderwijk (RGO) for helping us extract data from
the archive,
the Carlsberg Meridian Group (RGO) for extinction data, 
and Javier Diaz (IAC) for information about street-lighting
on La Palma.
SLE carried out part of 
this work while a summer student at the Isaac Newton Group
in 1996.

{\normalsize\bf References}\\
Benn, C. R., \& Ellison S. L., 1998, La Palma Technical Note 115\\
Garstang R.H., 1989, PASP, 101, 306\\
Hunter T.B. \& Crawford D.L., 1991 in `Light Pollution, Radio Interference and
               Space Debris', ed. D.L. Crawford (PASP conference vol. 17),
               p.89 \\
Krisciunas K., 1997, PASP, 109, 1181 \\
Massey P., Gronwall C., Pilachowski C., 1990, PASP, 102, 1046 \\
Roach F.E. \& Gordon J.L., 1973, `The Light of the Night Sky' (Dordrecht: Reidel)\\
Smith F.G., 1979, Trans IAU 17A, 220 \\
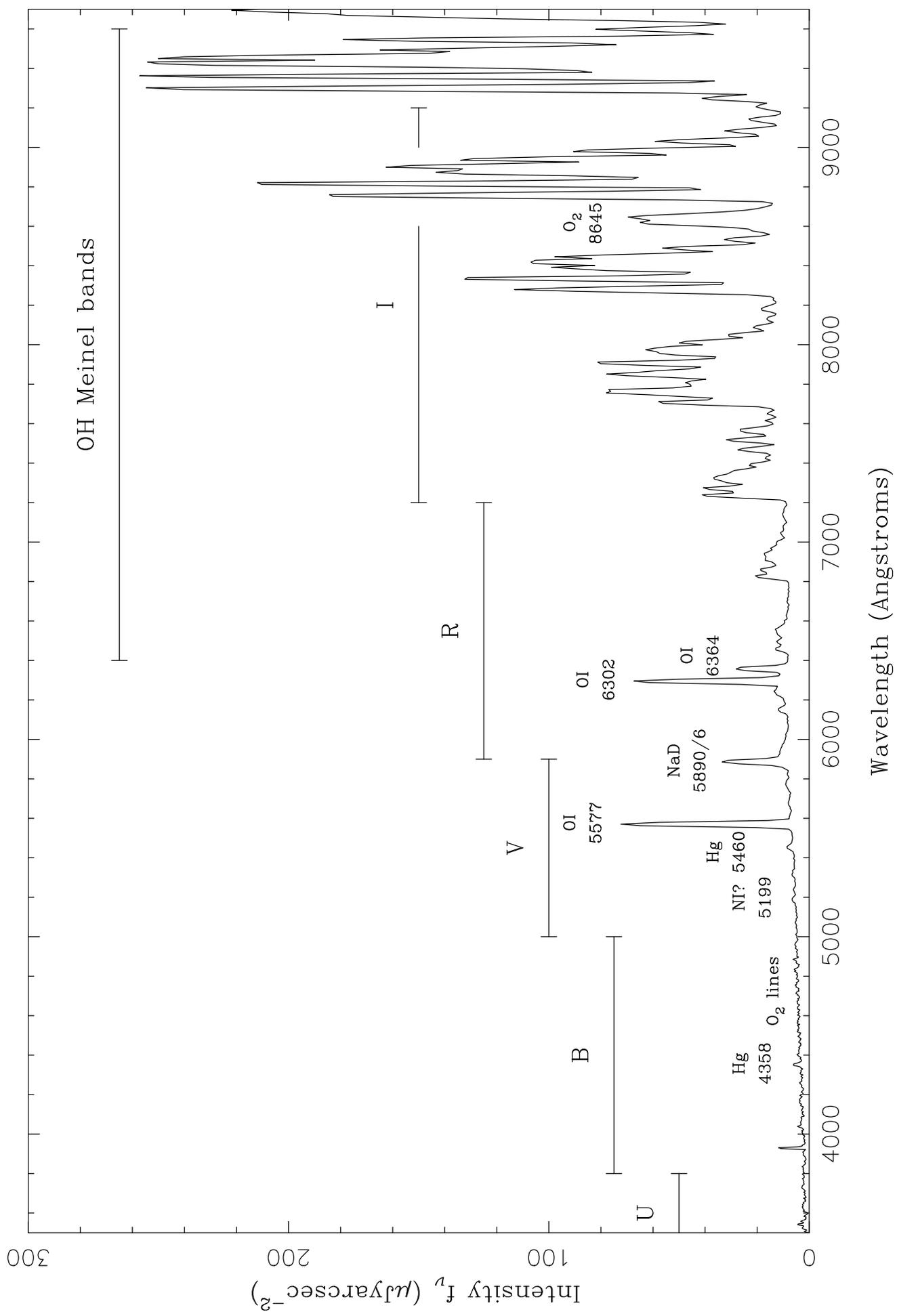
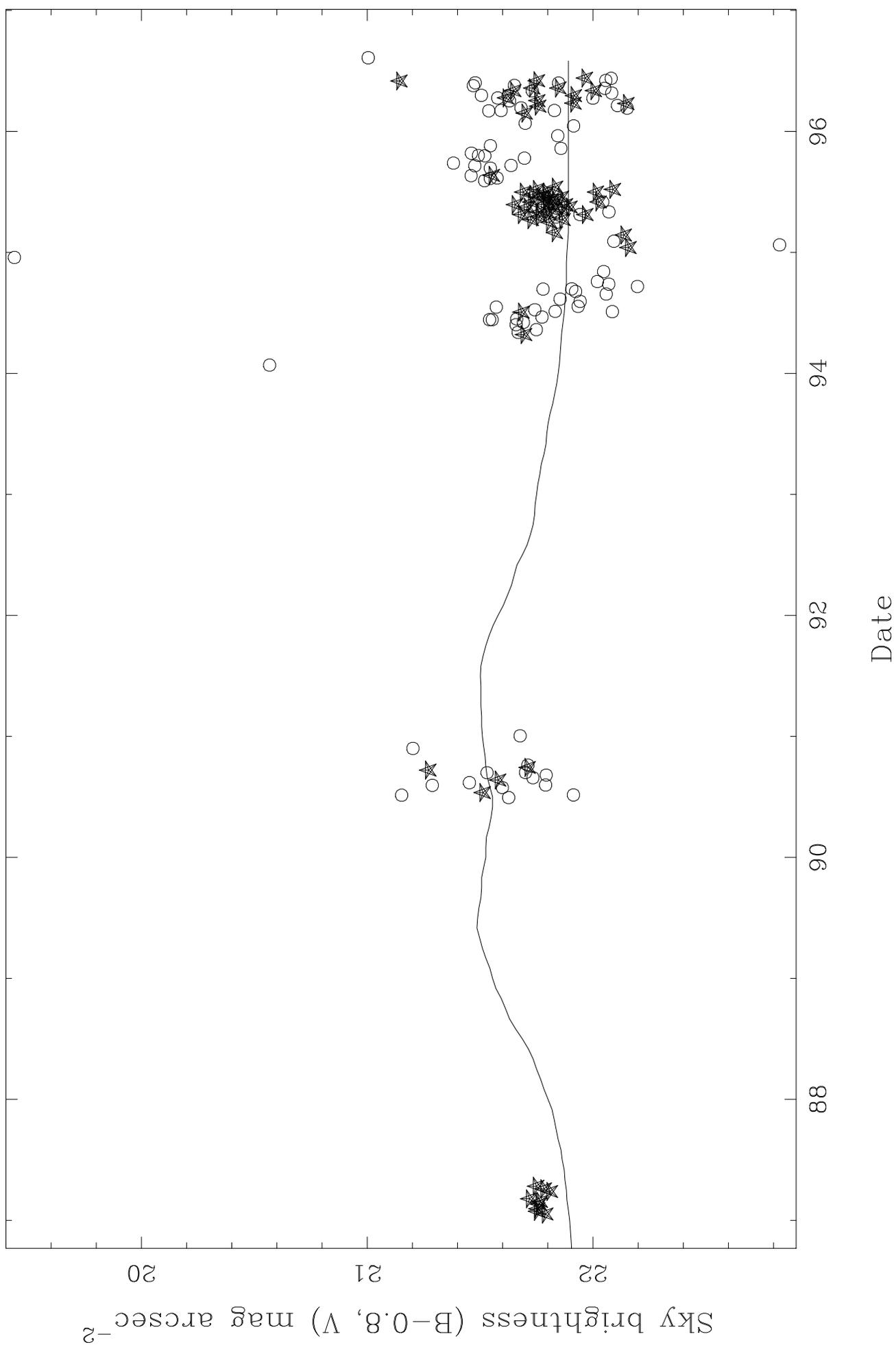
\end{document}